\documentclass[a4paper]{aa}
\usepackage{psfig}
\usepackage{aabib} 

\begin{document}

\title{{\em XMM-Newton} survey of the low-metallicity open cluster NGC~2516}

\author{S. Sciortino\inst{1} \and G. Micela\inst{1} 
\and F. Damiani\inst{1} \and E. Flaccomio\inst{1}
\and K. Briggs\inst{2} \and M. Denby\inst{2} \and J. Pye\inst{2} 
\and \\ N. Grosso\inst{3} \and A. M. Read\inst{3} 
\and P. Gondoin\inst{4} \and R. D. Jeffries\inst{5}}

\institute{Osservatorio Astronomico di Palermo G.S. Vaiana,
 Piazza del Parlamento 1, I-90134 Palermo, Italy
\and                                              
X-ray Astronomy Group, Department of Physics and Astronomy, University of
Leicester, Leicester, LE1 7RH, UK
\and
Max Plank Institut f\"{u}r Extraterrestrische Physik,
Giessenbachstra\ss{e}, D-85741 Garching bei M\"{u}nchen, Germany
\and
European Space Research and Technology Center, 2200 AG Noordwijk ZH, 
The Netherlands
\and
Department of Physics and Astronomy, Keele University, Staffordshire, ST5 5BG, UK}

\offprints{S. Sciortino (sciorti@oapa.astropa.unipa.it)}

\date{Received 30 September 2000 / Accepted 4 November 2000 }

\abstract{
We present the first results of an {\em XMM-Newton EPIC} survey of NGC
2516, a southern low-metallicity open cluster with an age close to the
Pleiades.  The attained limiting sensitivity is of $\sim 2.4\times
10^{-15}$ erg~sec$^{-1}$~cm$^{-2}$ in the 0.1--4.0 keV bandpass.  This
has been achieved by summing the data of the MOS and PN cameras of two
distinct observations for a total exposure time of $\sim$ 33 ks and by
analyzing the summed data set with the wavelet detection algorithm
developed at Osservatorio Astronomico di Palermo (OAPA) that has
yielded over 200 X-ray detections. Using data just from a single
exposure or from a single camera would have reduced by a factor 2--4
our limiting sensitivity and would have resulted in 25--40\% less X-ray
detections.  To date, 129 detections have as counterparts one or more
of the 540 photometrically selected cluster members in the surveyed
region, for a total of 147 likely detected members, with unique
identification in 112 cases.  We derive the X-ray luminosity functions
(XLF) of NGC~2516 members of different spectral types and compare them
with those of the more metal rich, approximately coeval Pleiades
cluster, finding the NGC~2516 photometrically selected dG and dK stars
less luminous than the Pleiades. The XLFs of the NGC~2516 and of the
Pleiades dM stars are indistinguishable.  We compare the {\em
XMM-Newton} results with those recently obtained with {\em Chandra}.
 }

\maketitle                                                

\section{Introduction}

NGC~2516 has been chosen as an {\em XMM-Newton} calibration target in order to
``bore sight'' the alignment of the X-ray telescope as well as
the relative positions of the EPIC cameras CCDs since it is
a region rich of X-ray emitting young stars. 
Apart from this role in the early phase of the 
{\em XMM-Newton} mission, NGC~2516
plays an important role in the study of stellar physics. It is a southern, 
relatively
nearby ($\sim$ 387 pc; \cite{JTP97}), open cluster with metallicity below solar 
([Fe/H]~=~ $-$0.32$\pm$0.06, \cite{JTP97}; $-$0.42, \cite{Cam85}; $-$0.18$\pm$0.08,
\cite{JJT98}), affected by moderate extinction
($E(B-V)~=~0.12$), and an age ($\sim$ 10$^8$ yr) similar to that of the Pleiades. 

Because of its metallicity, it plays a crucial role in exploring the
relation between stellar structure, metallicity, and coronal emission
level. Such a relation is expected to exist since metallicity
affects the depth of the convection zone that in turn should affect the
efficiency of dynamo action, of which X-ray luminosity is a proxy (cf.
discussion in Micela et al., 2000). Indeed Jeffries et al. (1997),
based on {\em ROSAT-PSPC} data, have found that the brightest X-ray luminous dG
NGC~2516 members are less luminous than the brightest X-ray luminous
solar-metallicity, similar age, dG in the Pleiades. More recently
Harnden et al. (2000), based on {\em Chandra} data, have found that the
median X-ray luminosity of NGC~2516 dF stars is higher than that of the
Pleiades, while there is an indication of the NGC~2516 dG and dK being
less X-ray luminous than the Pleiades (albeit this result can be
subject to uncertainties in assessing member status).  The availability
of deep {\em XMM-Newton} data allow us to further investigate this
matter.

Our paper is organized as follows: in \S 2 we will describe the X-ray
data and analysis, in \S 3 we will describe the results we have obtained and in
\S 4 we will summarize and discuss our findings.

\section{X-ray Observations and data analysis}

\begin{figure*}
\vbox{
\centerline{\psfig{file=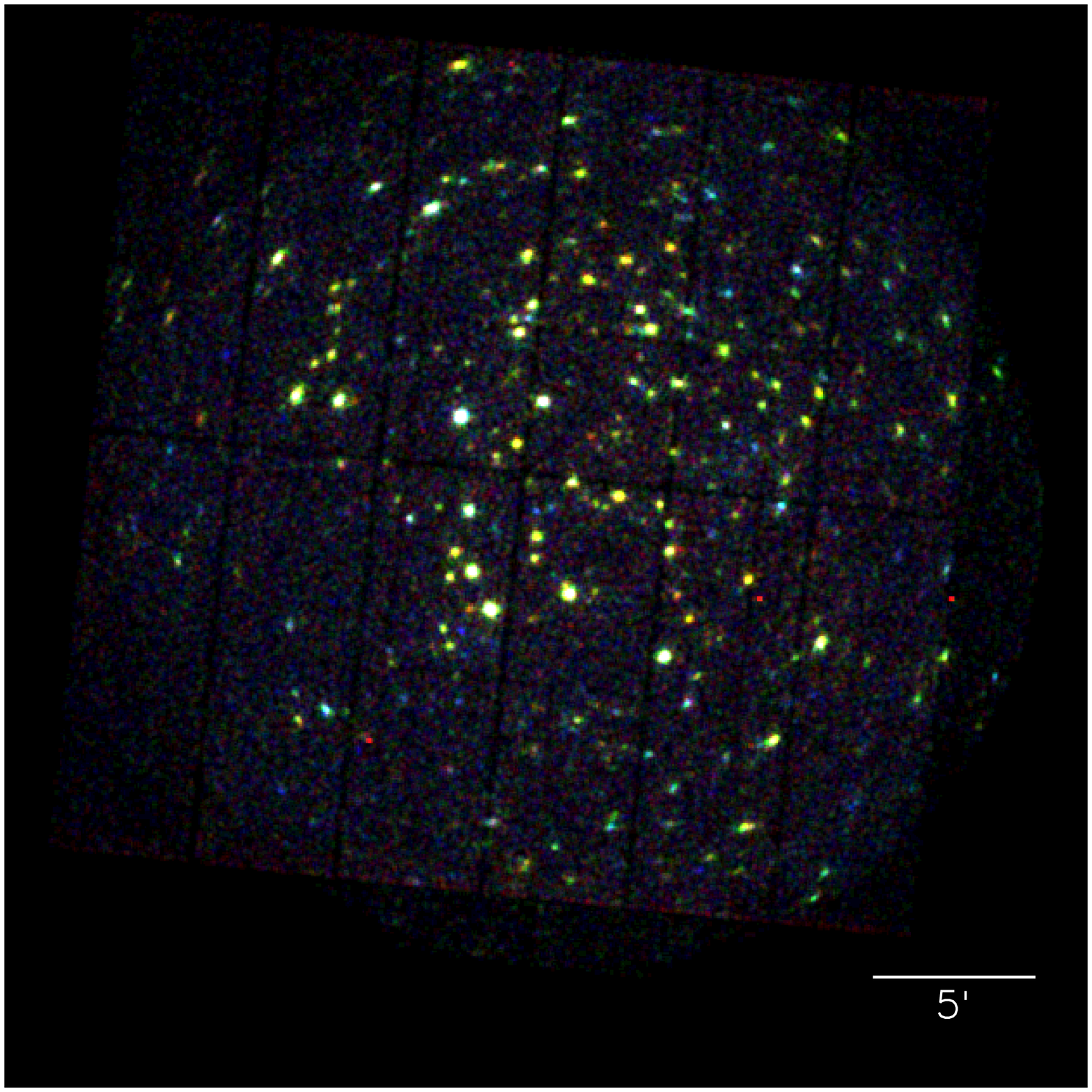,width=13.0cm}}
\vspace{0.6cm}}
\caption{ {\em XMM-Newton} EPIC ``spectral'' image for the summed B 
and C observations (N up, E to left). Pixel size is of 3.2$^{\prime\prime}$ on a side.
The image has been color-coded by assigning three distinct colors to
three distinct energy bands, namely: 0.3--0.8 keV (red), 0.8--1.5 keV (green)
and 1.5--4.0 keV (blue) and has been smoothed using a moving box of 3\,$\times$ \,3 pixels.
The bar at the bottom right indicates the image scale. Note the presence of
a handful of markedly soft sources, likely foreground soft stellar sources.
}
\label{fig:all.image}
\end{figure*}

\begin{table*}[t]
\caption{Summary of the NGC~2516 {\em XMM-Newton} Calibration Data until July 1-st 2000}
\begin{tabular} {c c c c c c} \hline
Obs ID & RA (h m s) & Dec (d m s) & Exp. Time$^{a}$ & Clean Exp. Time$^{b}$ & Obs Date \\ \hline
A & 7 58 33 & $-$60 52 54 & 7.7 ks & 7.2 ks & 12 March 2000 \\
B & 7 58 29 & $-$60 53 32 & 20.0 ks & 18.6 ks & 6 April 2000 \\
C & 7 58 29 & $-$60 53 32 & 17.0 ks & 15.1 ks & 7 April 2000 \\
D & 7 58 18 & $-$60 47 13 & 29.2 ks & 22.8 ks & 10 June 2000 \\ \hline
\end{tabular}

\noindent
a - Exposure times are for the PN camera. The MOS camera ones usually
differ only by a few hundreds of seconds, except for observation D for
which MOS exposure times are $\sim$ 2 ks shorter.
b - Cleaned exposure times are essentially the same for MOS and PN cameras.
\label{tab:sumdata}
\end{table*}

During the calibration phase, lasted until July 1-st 2000,
NGC~2516 has been observed 4 times (cf. Table \ref{tab:sumdata}) with
the {\em XMM-Newton} EPIC constituted of 2 MOS (\cite{T++00}) and 1
PN (\cite{S++00}) cameras.  In all cases observations have been
performed with the thick filter in the optical path.
The four data sets differ not
only for exposure times, but also for the
influence of random rapid increases in the background level
that occurs with erratic durations
as result of intense flux of soft protons in the earth's
magnetosphere, pseudo-focussed by the mirror system ({cf. \cite{J++00};
\cite{T++00}; and \cite{S++00}). Since those episodes dominate the
overall background level, in order to achieve the deepest limiting sensitivity we
have screened out those offending time segments. The resulting
``cleaned'' exposure times are summarized in Table \ref{tab:sumdata}. As a
matter of fact, for these four observations, we have to reject only a
small fraction of the exposure times, but with a substantial reduction
in the background level.  Outside of these episodes we have observed a
quite stable background level. 
Data have also been cleaned to remove electronic noise
that affects the MOS cameras and defects in CCD chips, such as dead or hot
pixels. Most of the data-cleaning procedure is currently performed by
the Standard Analysis System (SAS) reduction chains (procedures {\it
emchain} and {\it  epchain}), but a few hot pixels in the MOS cameras and
one hot column in the PN camera have been removed
afterwards. We have also rejected those events that because 
of their characteristics are unlikely to be bona-fide X-rays\footnote{
As a technical detail we have retained 
only those events whose {\it PATTERN} is $\leq$ 12 and
{\it FLAG} is $\leq$ 10.}.

We have analyzed the observations B and C that are pointed toward the
same direction.  As a first step we have considered 
the data of the
three EPIC cameras, 2 MOS (\cite{T++00}) and 1 PN (\cite{S++00}) and of
the B and C observations, separately. By cross-matching the individual
source lists we have been able to derive: i)
the relative spatial registering of the two MOS cameras (their mean
relative alignment is currently of $\sim$ 1$^{\prime\prime}$ 
with a rms of 3$^{\prime\prime}$) as well
as that of the PN one; ii) a preliminary absolute spatial registering of the
MOS cameras that is currently accurate to $\sim$ 3$^{\prime\prime}$ as measured by
the median offset between X-ray and optical positions. These latter are taken
from the catalogue we have used for identification purpose. By cross-matching 
the positions of
the more than 100 sources detected by using the SAS task {\it
edetect\_chain} on the individual MOS and PN data sets of the
observation B we have found that at the time this work was done (end of
Sept. 2000) an offset and a rotation are required to reconcile the PN
and MOS astrometry; the major effect being a shift of the MOS derived
positions of about 10$^{\prime\prime}$ north of the PN ones.  
We have verified that the derived
correction parameters work well also for observation C.  As we
have discussed (cf. also Fig.~\ref{DSS.image} below), the MOS
astrometry is in reasonable agreement with optical positions, hence we have
``moved'' the PN data to the MOS astrometric system and have summed in a
single data set the 6 distinct MOS and PN data sets of the observations
B and C. 

Since our aim was to reach the deepest limiting sensitivity
and we are interested in the study of coronal emission from normal
stars we have further restricted our following analysis to photons with
energy in the 0.3--4.0 keV bandpass\footnote{Events with energy below
0.3 keV are largely unrelated to bona-fide X-rays.}.  With this choice
we have reduced the number of image counts (largely dominated by
background events in our case) from $\sim$ 376.0 to 259.5 kcnt, i.e. we
have effectively reduced the background level by about 45\%.  The
resulting summed image is shown in Fig.~\ref{fig:all.image}, with
different color coding for the spectral distribution of collected counts in
each image pixel.  Summed data have then been searched for
deriving the final list of sources (see below).

\begin{figure}
\centerline{\psfig{figure=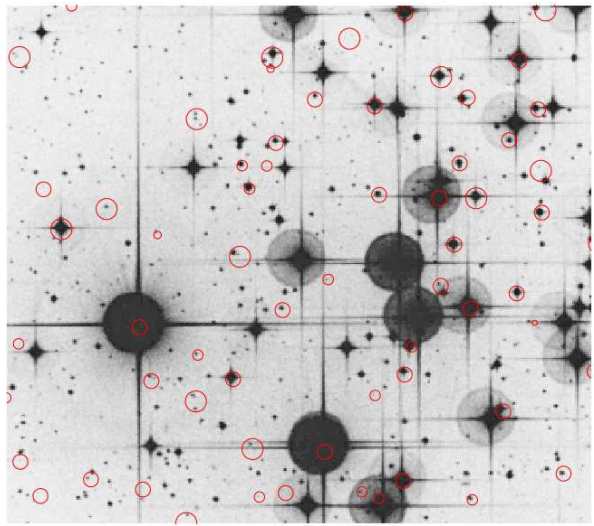,width=7.8cm}}
\caption{DSS image (N up, E to left) of the central part 
($\sim$ 10$^{\prime}$.5\, $\times$ \,10$^{\prime}$) 
of the surveyed region with overlayed red circles at source positions. 
Circle size is equal to the scale at which each source has been 
detected with the highest significance. Note the good agreement between optical
and X-ray positions.}
\label{DSS.image}
\end{figure}

It is worth mentioning that the data have been analyzed without
including the spacecraft history file (i.e. no detailed aspect
information was available to us at the time this work has been done)
and the pointing of the telescope was assumed to be stable at the
nominal value. The quality of the resulting summed image makes us
confident that fluctuations in spacecraft pointing have introduced only
minor effects.

At the time this work has been done SAS had not full provision
for the analysis of summed data sets, so we have to use
tools coming from various packages (SAS, IRAF, FTOOLS, IDL) as well as some
ad-hoc developed code. In particular, source detection on the
summed data set has been performed with a wavelet transform (WT) detection
algorithm developed at Osservatorio Astronomico di Palermo 
specifically for the {\em XMM-Newton} EPIC
data.  The basic features of this code are inherited from the
{\em ROSAT/PSPC} version (Damiani et al.\ 1997a,b).  Briefly, it performs a
multi-scale analysis of the data, efficiently detecting sources ranging
in size from point-like to a few
arcminutes. The newest features of the {\em XMM-Newton} WT code
include: the operation on images with pixel size of 2$^{\prime\prime}$, 
the full scan
of the entire image at all explored scales to take advantage of the
almost uniform width (HEW $\sim$ 15$^{\prime\prime}$) of the {\em XMM-Newton} PSF (cf.
\cite{J++00}).  Further details on this algorithm will be presented by
Damiani et al. (in preparation). In the present application we
have searched the image at four distinct scales from 4$^{\prime\prime}$ 
to 11$^{\prime\prime}$.3,
more than adequate for detecting point-like sources.  In order to
derive the appropriate detection threshold that limits spurious
detections to the desired number (chosen here as 1 per field) we have
performed a large set of simulations of pure-background EPIC images
{\it with the same exposure map and the same background level} as the
actual data set we have analyzed.  We have found 208 sources (with one
predicted spurious) and have measured their net counts. Sources as
faint as 25--30 counts have been detected.

Given the assumption that we have made on pointing stability, it is quite
reassuring that the overall attained positional accuracy is quite good
and is within expectations.  This is well illustrated in Fig.~\ref{DSS.image} 
where we have overlayed circles centered at the X-ray source positions
on the DSS image of the central part of the surveyed region.

\subsection{X-ray flux and luminosity determination}

In order to derive source count rates we have computed (using SAS)
the exposure maps for
the 6 (3 cameras $\times$ 2 observations) distinct data sets we have
used deriving the total exposure map that we have adopted 
in the rest of our analysis.  We have to remember that the MOS
(\cite{T++00}) and PN (\cite{S++00}) cameras have distinctly different
effective areas, or in other words the number of counts collected for a
given source is different in the MOS and PN cameras.  In order to
circumvent this difficulty we have computed from the actual data the
ratio of MOS1, MOS2 and PN source counts.
The median of the ratio of PN and single-MOS collected counts, in the 0.3--4.0 keV band,
is $\sim$ 2.5, i.e. on average the PN camera collects 2.5 times more counts than
each single MOS camera.  This factor has been included in the
derivation of the summed exposure map.  We recognize that a) this
factor could be a function of source intrinsic spectrum and b) its
determination is subject to uncertainties since measured ratios have
a somewhat large scatter.  In deriving X-ray flux we have assumed a
conversion factor from counts to intrinsic source flux in the 0.1--4.0
keV bandpass\footnote{We have chosen this bandpass for easy comparison
with published {\em ROSAT} and {\em Chandra} results.},
as computed (using PIMMS) for a Raymond-Smith spectrum with $kT$=0.54 keV
and $N_H = 7.5\times 10^{20}$ cm$^{-2}$, of $7.1\times 10^{-12}$
{\rm erg~cm$^{-2}$~cnt$^{-1}$} for 
a single MOS camera. We note that this broad band conversion factor is
rather insensitive to metallicity variations up to a factor 3.
With the above conversion factor, the net exposure time of 
$\sim$ 33 ks of the summed
B+C data set, translates to an equivalent $\sim$ 150 ks of a single MOS
camera (or to an equivalent $\sim$ 60 ks of the PN camera). 

\begin{figure}
\centerline{\psfig{figure=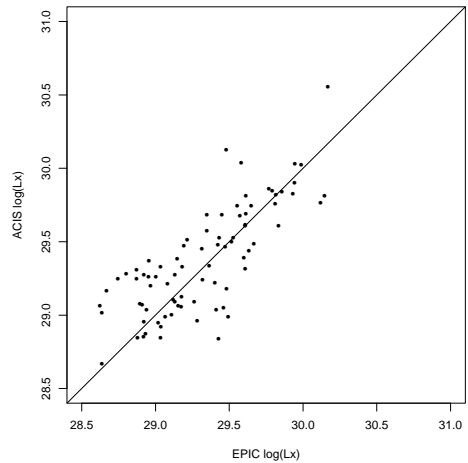,width=6.5cm}}
\caption{Comparison of the {\em XMM-Newton} EPIC and {\em Chandra} ACIS-I 
X-ray luminosities (in the 0.1--4.0 keV
bandpass) of detected NGC~2516 members.  The solid line represents the locus of
equal luminosities.}
\label{fig:Lx_Lx}
\end{figure}

The single-MOS equivalent rate of weaker sources is $\sim 3.3\times 10^{-4}$
cnt~sec$^{-1}$ (indeed we have detected only three sources below this
value), that corresponds to $\sim 2.35\times 10^{-15}$ erg~sec$^{-1}$~
cm$^{-2}$, well below the limiting sensitivity ($\sim 6\times 10^{-15}$ 
erg~sec$^{-1}$~cm$^{-2}$), achieved by the 60 ks {\em ROSAT-PSPC} 
observation analyzed
by Jeffries et al. (1997), and with a spatial resolution (EPIC FWHM
$\sim$ 6$^{\prime\prime}$) better than the PSPC whose PSF had a FWHM of
$\sim$ 20$^{\prime\prime}$. Summing all the available data allows us to
reach the deepest possible limiting sensitivity. With the
data taken just with observation B we would have reached a limiting
sensitivity of $\sim 8\times 10^{-15}$ {\rm erg~sec$^{-1}$~cm$^{-2}$}, and
$\sim 3.5\times 10^{-15}$ {\rm erg~sec$^{-1}$~cm$^{-2}$} for a single MOS
and the PN cameras, respectively. Summing up the data of all cameras for
the observation B we would have reached a limiting sensitivity of
$\sim 3\times 10^{-15}$ {\rm erg~sec$^{-1}$~cm$^{-2}$}, detecting only
about 160 sources. From the above numbers, is is also clear that,
for exposure greater than $\sim$ 20 ks EPIC enters in the 
background-limited regime and the limiting sensitivity does not scale 
linearly with exposure time\footnote{Thanks to the sharp
{\em Chandra} PSF an ACIS-I observation
starts to become background-limited for exposure time greater than $\sim$ 200 ks.}.

The deepest (i.e. on-axis) limiting sensitivity of the present {\em
XMM-Newton} 33 ks observation is similar to that attained, again
on-axis, with a $\sim$ 20 ks
{\em Chandra} ACIS-I observation (\cite{H++00}), and it varies 
by a factor 2 within the central 11 arcmin region (to be compared with
a factor 5 variation in the case of {\em Chandra}); in other words the
present
{\em XMM-Newton} observation has a total grasp (for the FOV in common) that is 
$\sim$ 2 times better than the {\em Chandra} one. Since the present
observation has been done with the thick filter we expect that, under the assumption
of an approximately constant background level, the attainable limiting sensitivity (in the
somewhat soft band we have considered) could benefit from the use of the medium (or the
thin) filter.

In order to gain further confidence in the adopted procedure we have compared the {\em
XMM-Newton} EPIC X-ray luminosities with those derived from {\em Chandra}
observations (\cite{H++00}) for the members detected by both observatories. The resulting
scatter plot in Fig.~\ref{fig:Lx_Lx} shows a good agreement,
especially if one considers source variability, of which we have found clear 
evidence in the {\em XMM-Newton} data, as well as different source spectra.
As matter of fact we have detected 108 out of the 147 {\em Chandra} ACIS-I sources, and 
we have detected 34 new sources in the common FOV. A large fraction of the new
sources falls in the external part of the {\em Chandra} FOV. About half of the sources 
missed with EPIC have been detected in the central part of the {\em Chandra} FOV
corresponding to a somewhat external and less sensitive region of the EPIC FOV, the others
are likely to be variable. As stated above we have found clear evidence of variability
that will be discussed in a forthcoming paper.
Deeper limiting sensitivity 
will be reached by including in a future analysis the data from observations A and D
\footnote{Since we are already in the background-limited regime we estimate to
improve the limiting sensitivity by a factor $\sim$ $\sqrt 2$ for the
about twofold increases in the exposure time.}.

\section{Results}

\subsection{Cluster Members and Identifications}

\begin{table}[t]
\caption{Surveyed and detected NGC~2516 members vs. spectral type}
\begin{tabular}{lrrrrrr} \hline
Spectral Type:   &  B &  dA &  dF &  dG &  dK &  dM \\ \hline
\multicolumn{7}{l}{\it Entire XMM-Newton FOV } \\
Observed & 31 & 43 & 34 & 80 & 205 & 146 \\
Detected & 14 & 11 & 16 & 31 &  49 &  26 \\ 
Perc. Det. & 45\% & 26\% & 47\% & 39\% & 24\% & 18\% \\ \hline
\multicolumn{7}{l}{\it XMM-Newton and Chandra Survey common FOV} \\
Observed & 23 & 23 & 17 & 32 & 67 & 40  \\
Detected & 10 & 8 & 14 & 19 & 32 &  14  \\ 
Perc. Det. & 43\% & 35\% & 82\% & 59\% & 48\% & 35\% \\ \hline
\end{tabular}
\label{tab:stars}
\end{table}

\begin{figure}[h]
\vbox{
\centerline{\psfig{file=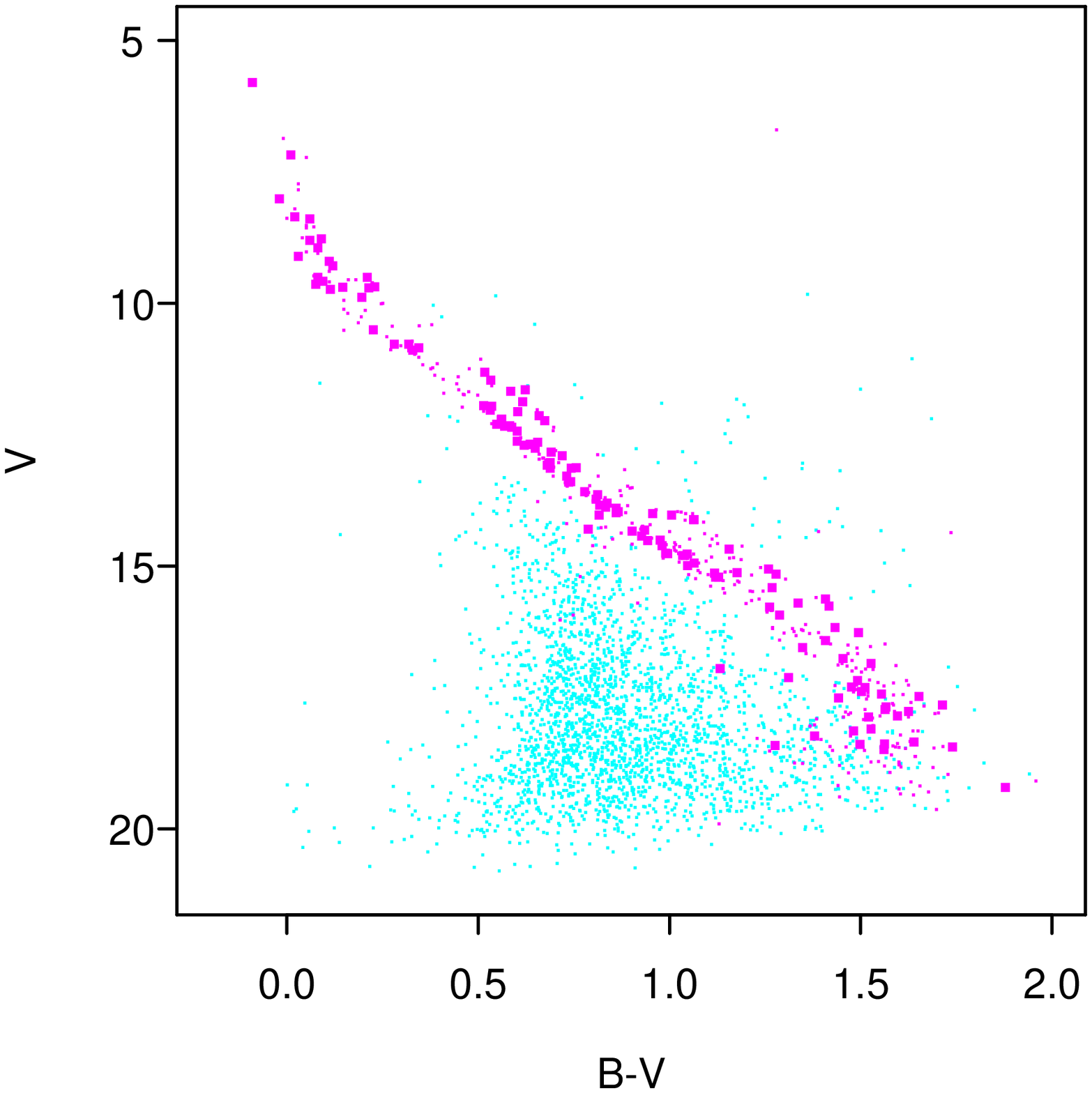,height=7.4cm}}
\vspace{0.2cm}
\centerline{\psfig{file=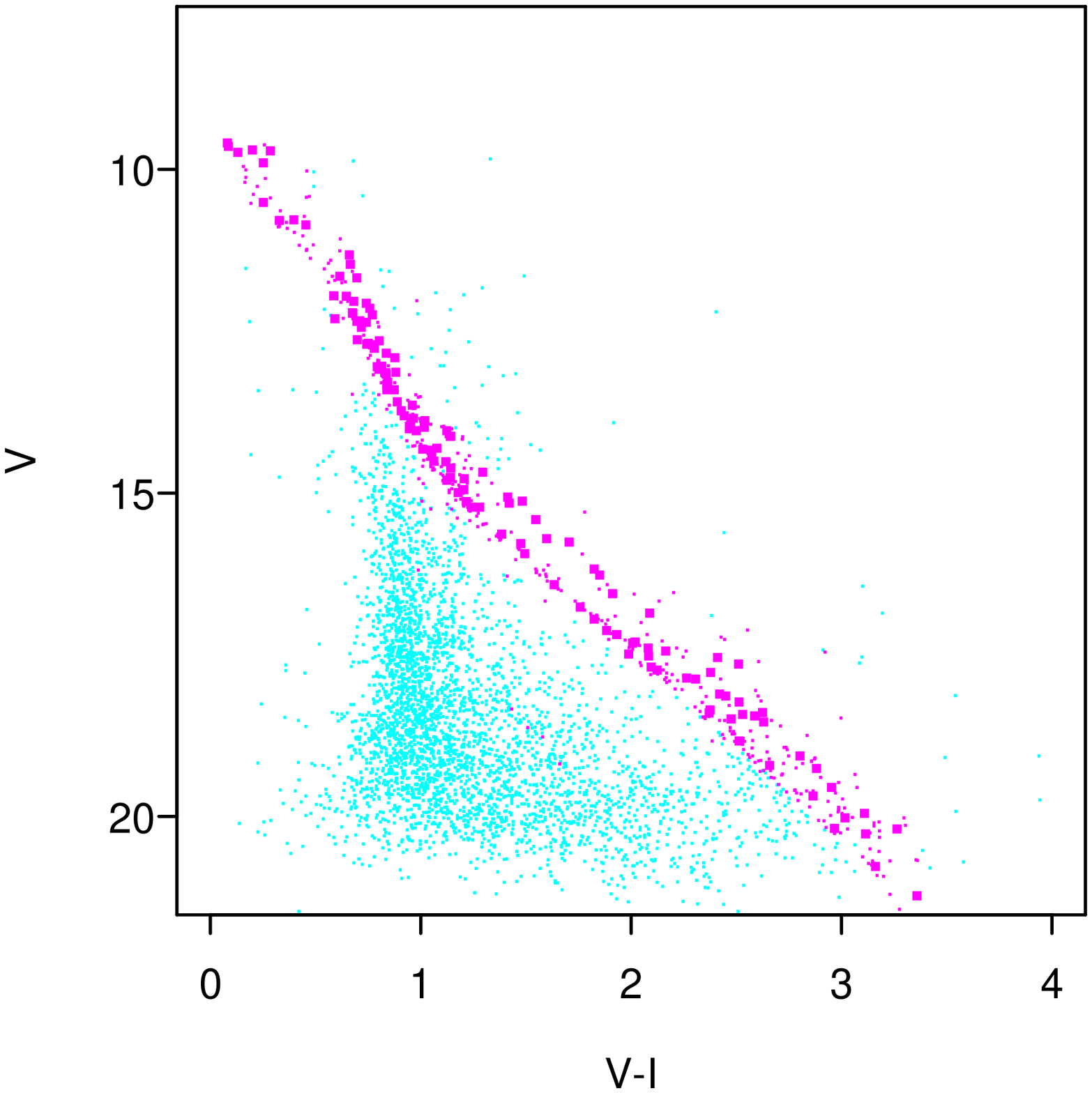,height=7.4cm}}}
\caption{(a) [top] $V$ vs. $(B-V)$ CMD of the objects in the surveyed region.
The non-members are indicated by blue dots, and a small fraction of them 
have been detected. The photometrically determined NGC~2516 cluster
members are indicated by red symbols: squares for detections and dots for 
members for which we derived only upper limits.  (b) [bottom] As above but 
for the $V$ vs. $(V-I)$ CMD.}
\label{fig:CMD}
\end{figure}               

\begin{table*}[t]
\caption{Comparison of our NGC~2516 and Pleiades (\protect\cite{MSH99}) median of $\log(L_{\rm X})$.
Values in parenthesis denote upper limits.}
\begin{tabular}{ccccc} \hline
Sp. & color & \multicolumn{2}{c}{NGC~2516} & Pleiades \\
 type    & range       & Ent. FOV & Restr. FOV    &          \\
\hline
B & $(B-V)_{\rm 0}$ $<$ 0            & 28.75  & 28.75 & ... \\
dA & 0 $<$ $(B-V)_{\rm 0}$ $<$ 0.3   & (28.75) & (28.75) & ... \\
dF & 0.3 $<$ $(B-V)_{\rm 0}$ $<$ 0.5 &  28.91  & 29.35 & 29.20 \\
dG & 0.5 $<$ $(B-V)_{\rm 0}$ $<$ 0.8 & (28.92) & 28.96 & 29.25 \\
dK & 0.93 $<$ $(V-I)_{\rm 0}$ $<$ 2.2 & (28.74) & 28.91 & 29.21 \\
dM & 2.2 $<$ $(V-I)_{\rm 0}$         & (28.67) & (28.67)  & 28.85 \\
\hline
\normalsize
\end{tabular}
\label{tab:XLF}
\end{table*} 

Color-magnitude diagrams (CMD) of the NGC~2516 photometrically
determined members in the present survey are shown in Fig.~\ref{fig:CMD}, 
together with all available photometric data of surveyed
non-members.  Photometry for the fainter stars is from a new 
catalogue (Jeffries, Thurston and Hambly, in preparation), while 
for stars brighter than $V \sim 9$ it is a collection of 
data coming from various sources (cf. \cite{JTP97}).
Cluster membership has been assigned by selecting stars within bands
around fiducial main sequences in HR diagrams of $V$ vs. $(B-V)$ and
$V$ vs. $(V-I)$, following a procedure described by Thurston (2000) and
reported by Harnden et al. (2000).
Our final list comprises those stars
fulfilling one of the above criteria\footnote{We have verified that 
adopting the more
restrictive selection in which both criteria have to be meet does not
change our results.}, when all three colors are available, and
the $V$ vs. $(V-I)$ criterion alone, for the reddest stars having no $(B-V)$
measurement. In summary, we have 540 likely photometric members out of
4117 objects with known photometry and falling in the {\em XMM-Newton}
field-of-view (FOV). Table \ref{tab:stars} gives the number of
members observed and detected (and the detected fraction) grouped
according to spectral types (B, dA, dF, dG , dK, and dM)
photometrically assigned on the basis of reddening-corrected
color ranges. One late-type giant falls in the FOV and has not
been detected. In the same table we report also the same quantities,
but for the restricted FOV in common with the {\em Chandra} survey
(\cite{H++00}).  
Identifications with X-ray sources 
have been performed through positional
matching with a match radius of 10$^{\prime\prime}$. We have chosen
this somewhat generous value to account for current limitation in the
absolute aspect reconstruction. Indeed with a match radius smaller than
8$^{\prime\prime}$ we would have lost bona-fide identifications.  Over
the entire FOV we have found 112 detections having a single cluster
member as counterpart, 16 detections having 2 members as counterparts
and 1 detection having 3 members as counterparts. The percentage of
chance coincidence between cluster members and X-ray sources is $\sim$
7\%, making us confident on the reliability of our identification
procedure.  The table shows also that in the entire {\em XMM-Newton}
FOV $\sim$24\% of surveyed cluster members have been detected,
while the detection fraction increases to 47\% in the smaller region in
common with {\em Chandra}. We will see in the following that
distinguishing between those two regions is important in interpreting
some of our results.

There are 79 (208$-$129) detections not associated with known cluster
members.  We have found at least one counterpart (in the
catalogue we have used) for other 39 sources, however we expect
$\sim$ 28 spurious identifications (by chance coincidence) among the
non-members. Since we have found
81 counterparts among the non-members, about 34\% of them are likely
to be spurious identifications.  
Many of these objects, as well as the non-identified
ones, are likely to be of extragalactic nature, a few others identified
with stars above the NGC~2516 main sequence, should be at smaller
distances than NGC~2516, and they are likely members of 
the young nearby stellar population present in all X-ray surveys (\cite{FBMS93}),
or given the pointing direction, members of the Gould Belt (\cite{GSS+98}).
Due to the ``fuzzy'' main-sequence definition at
faint magnitudes, a fraction of the sources identified with red stars
could be faint cluster members.

\subsection{X-ray Luminosity Functions}

\begin{figure*}
\vbox{
{\hbox{
\psfig{file=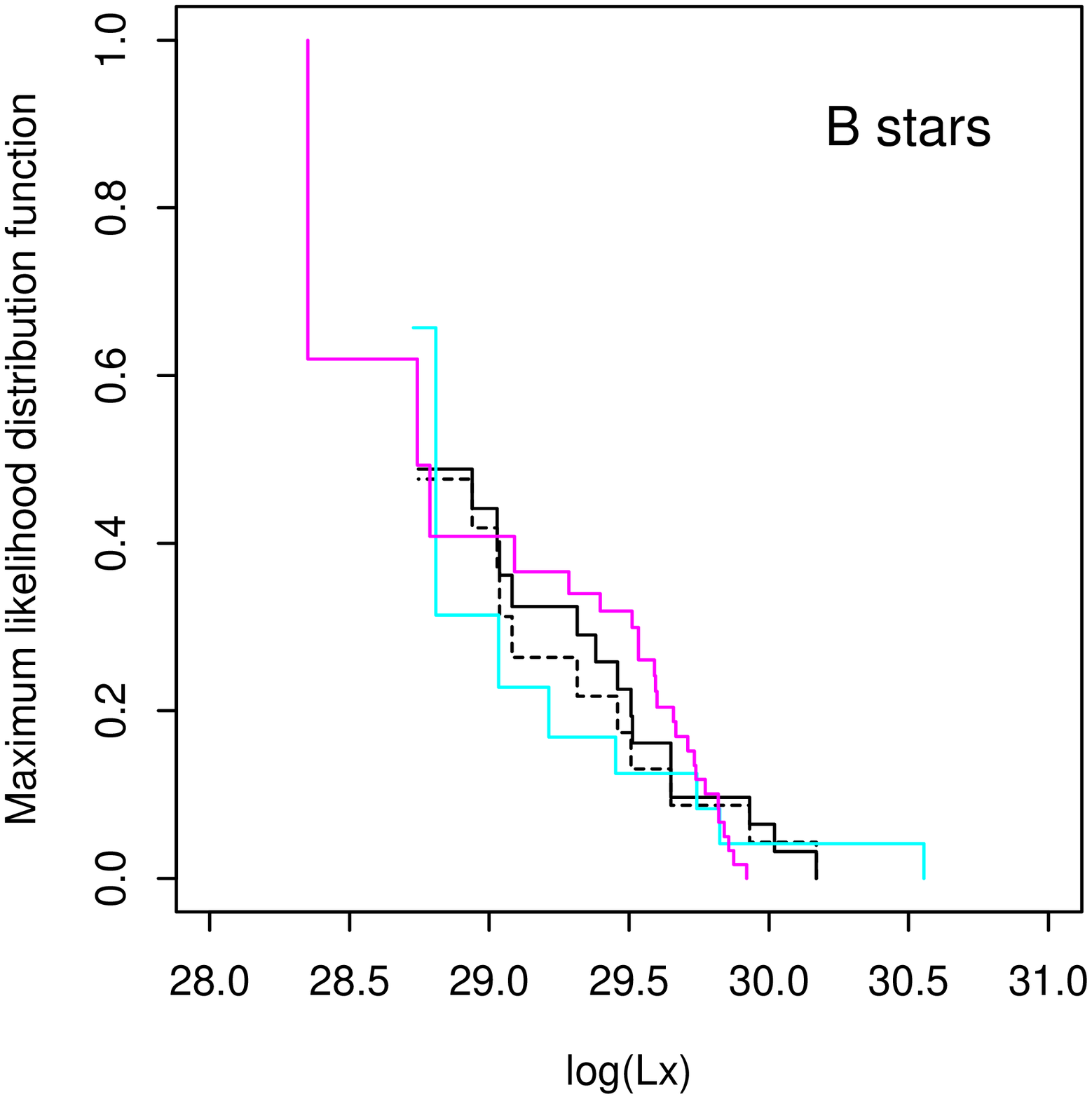,height=5.4cm}
\hspace{0.2cm}
\psfig{file=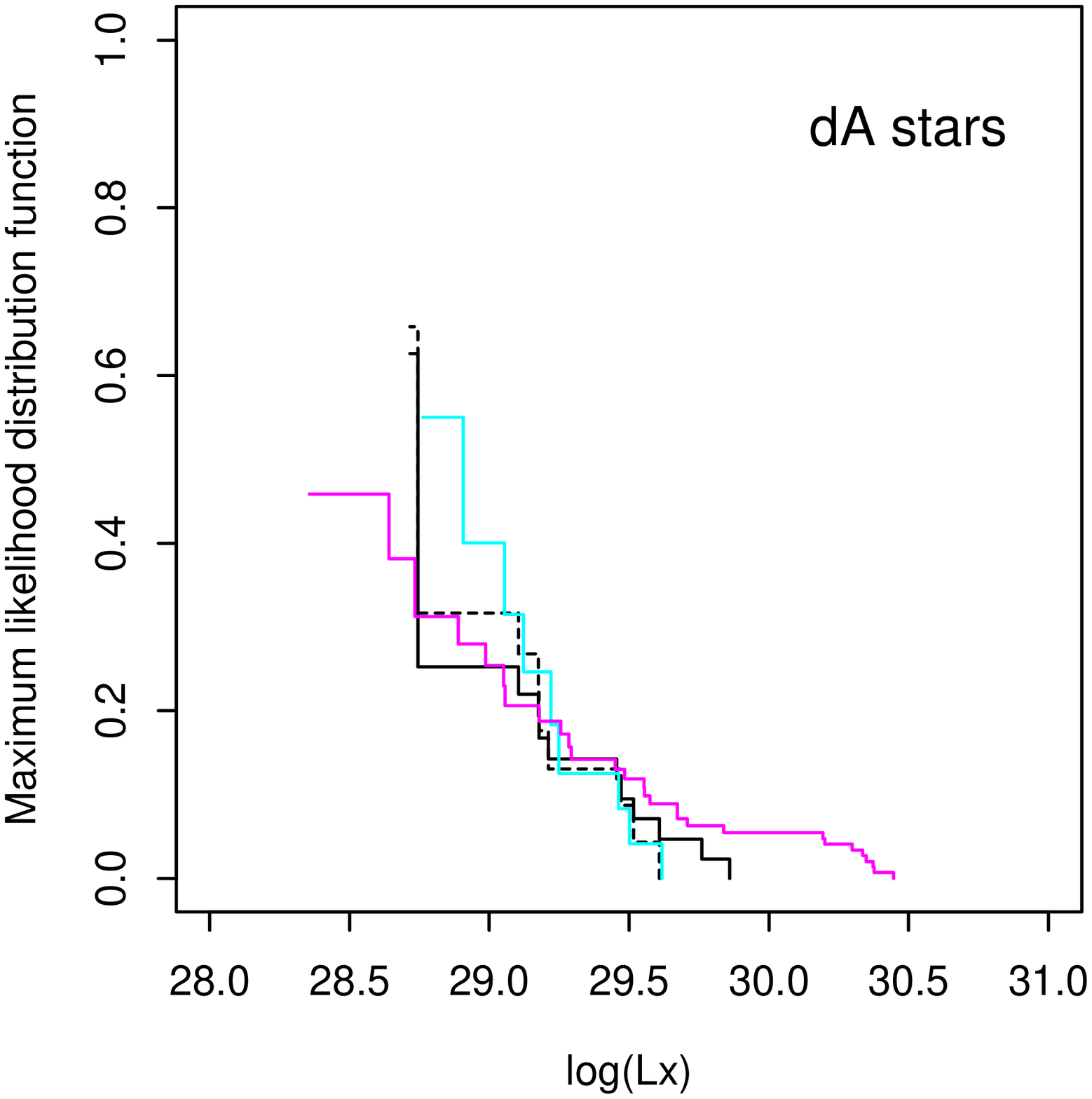,height=5.4cm}
\hspace{0.2cm}
\psfig{file=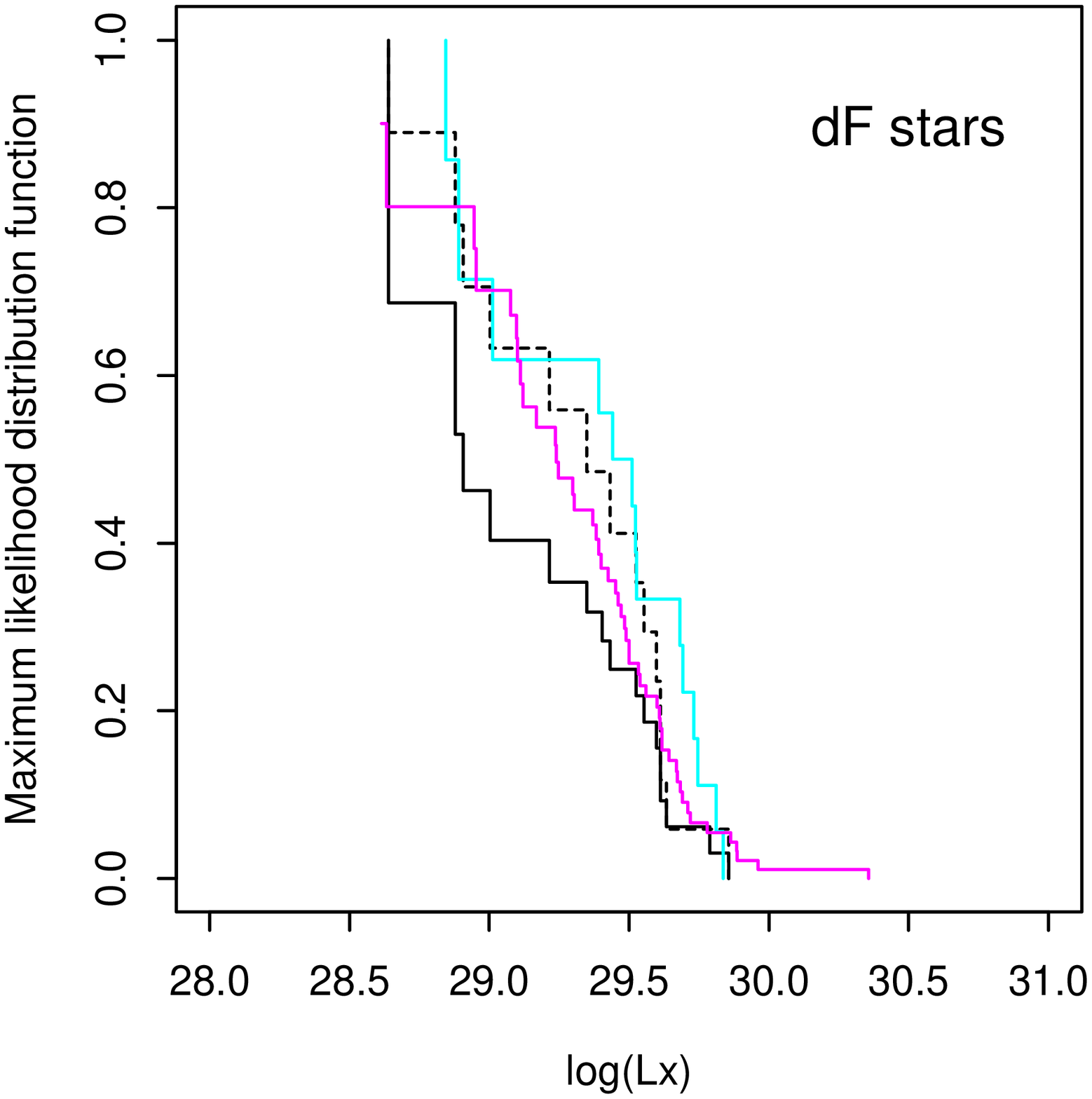,height=5.4cm}}}
\vspace{0.04cm}
{\hbox{
\psfig{file=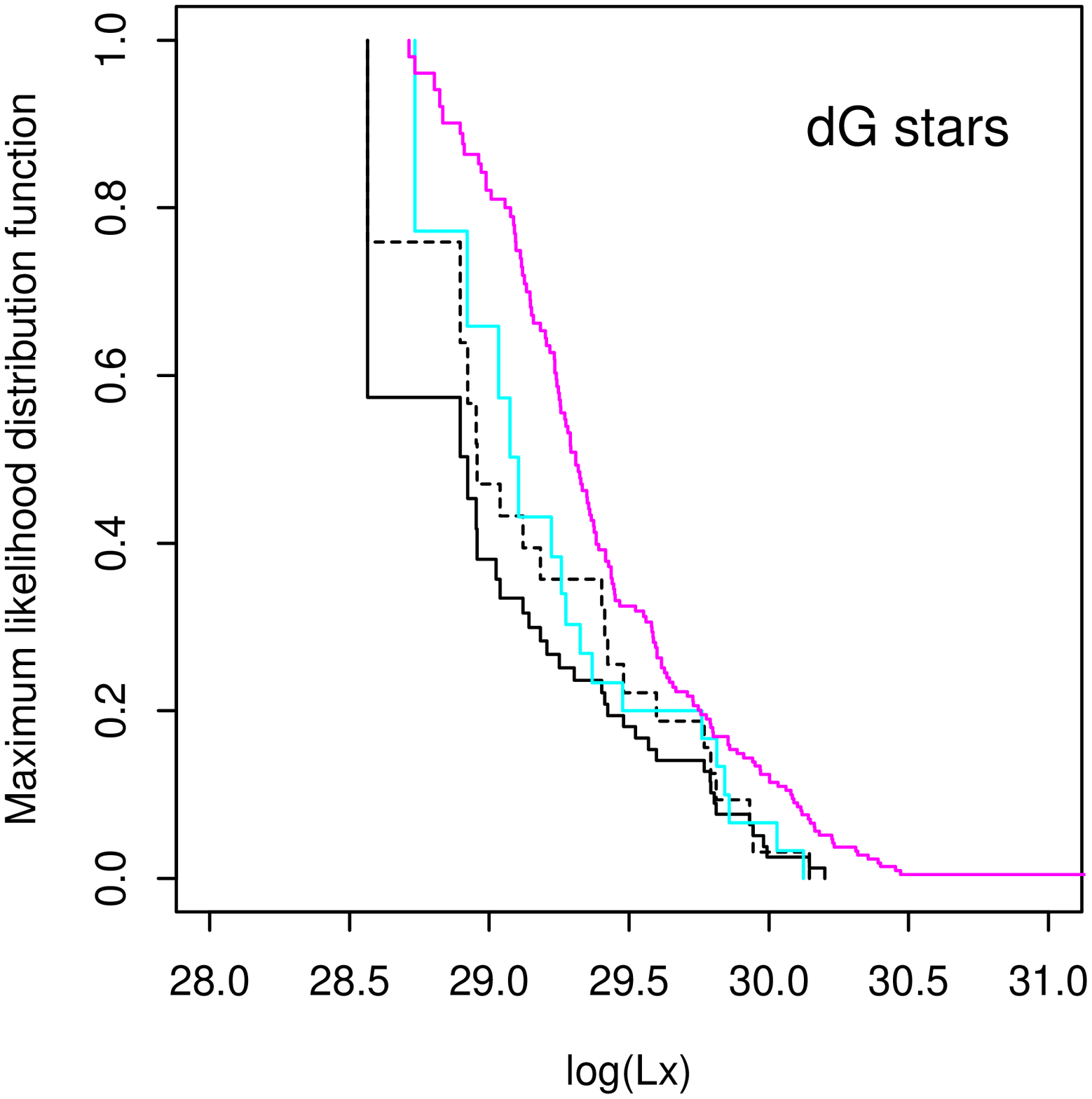,height=5.4cm}
\hspace{0.2cm}
\psfig{file=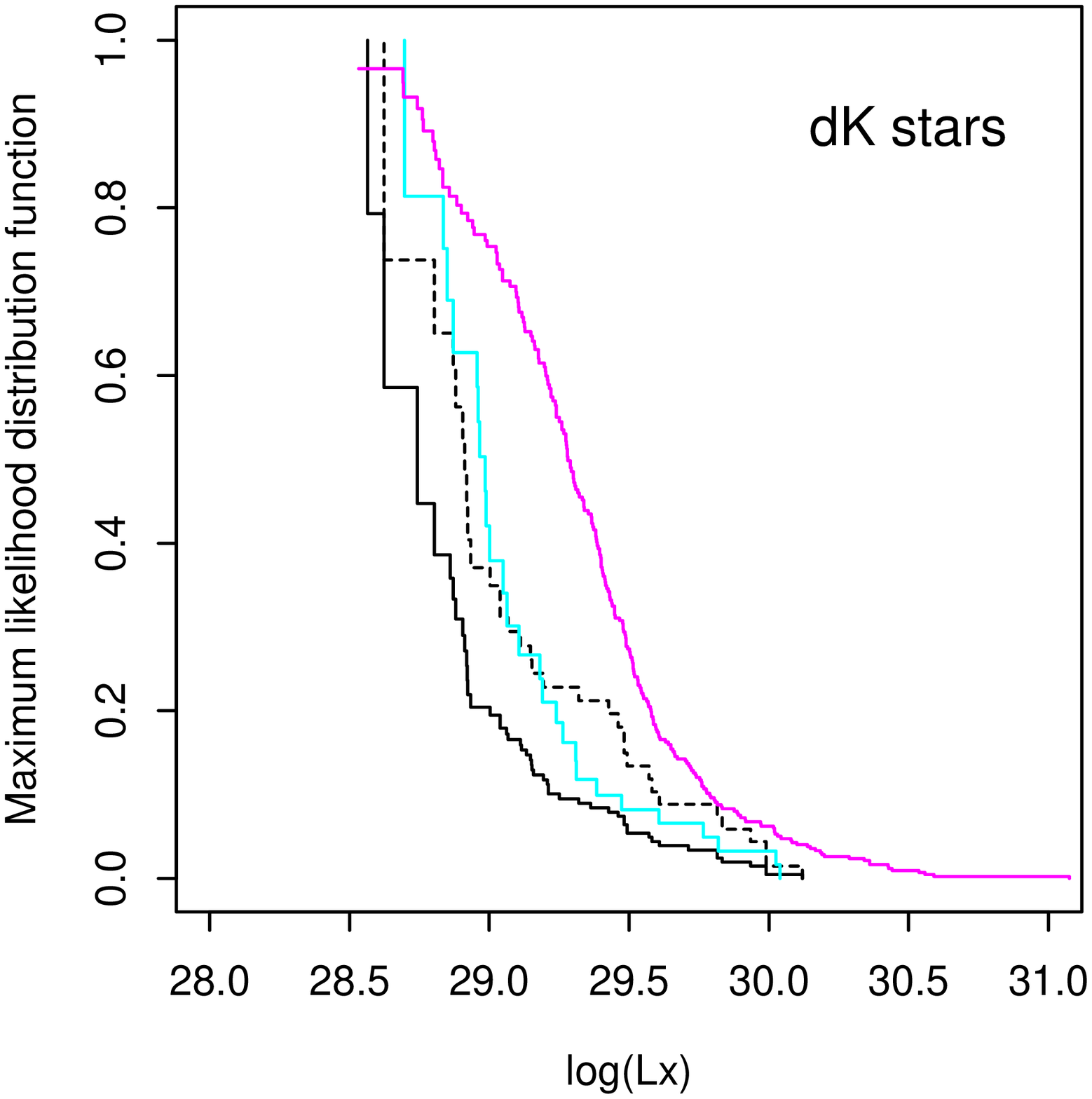,height=5.4cm}
\hspace{0.2cm}
\psfig{file=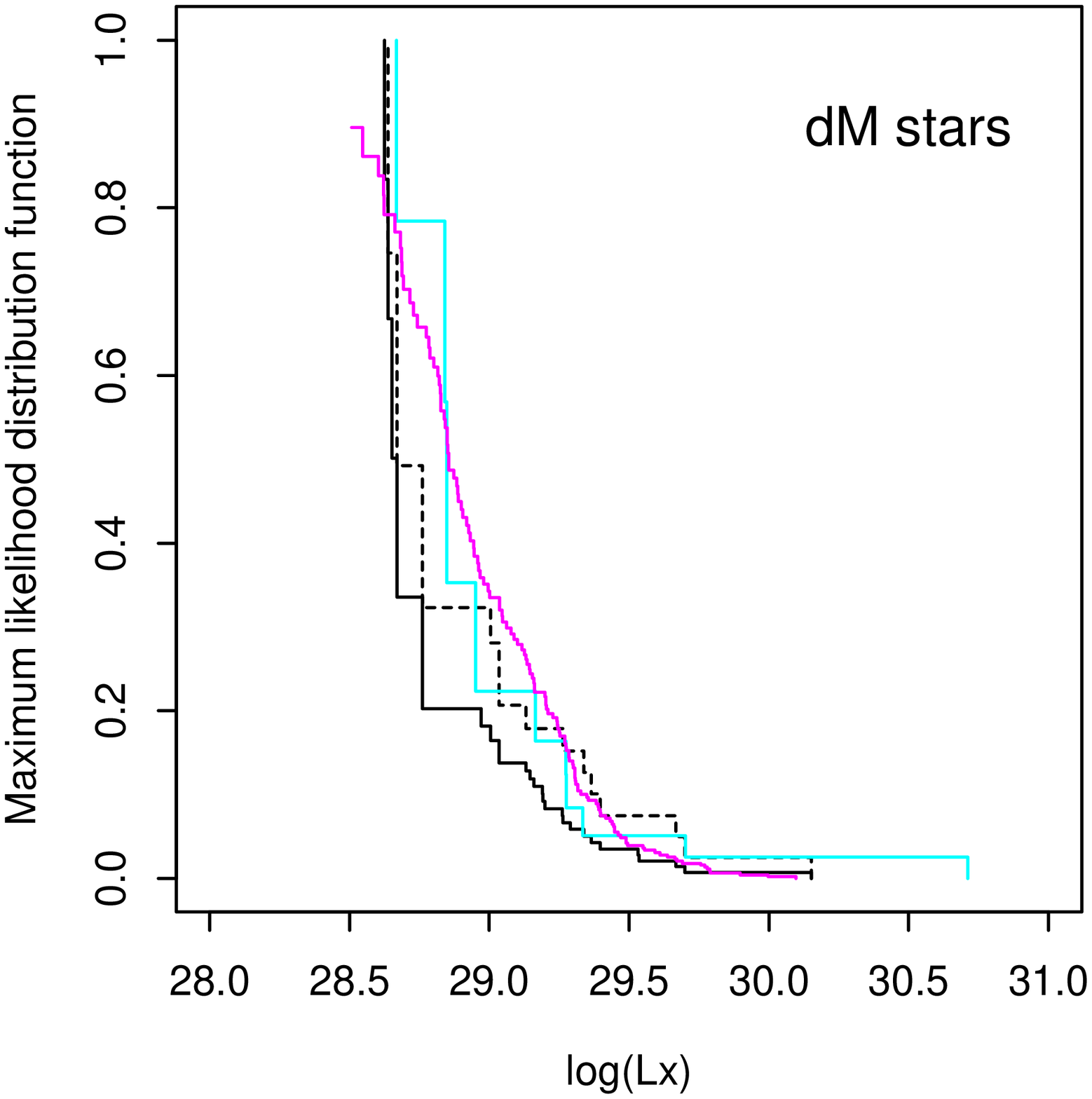,height=5.4cm}}}
}
\caption{{\em XMM-Newton} 0.1--4 keV X-ray luminosity
functions for NGC 2516 cluster members (black solid and dashed lines)
of spectral types B, dA, dF, dG, dK and dM as labeled. The solid lines
show the X-ray luminosity function for the entire EPIC FOV while the
dashed lines are for the sample of the stars falling in the {\em
Chandra} ACIS-I FOV. The analogous {\em Chandra} NGC~2516 X-ray luminosity 
functions (blue lines) and those for the Pleiades cluster (red lines),
derived from {\em ROSAT} data, are shown for comparison.}
\label{fig:XLF}
\end{figure*}

Including detections and upper-limits and making use of the
maximum-likelihood Kaplan-Meier estimator of integral distribution
functions in case of censored data (cf. \cite{JHS85}; \cite{FN85}) we have
computed the X-ray luminosity functions shown in Fig.~\ref{fig:XLF} for
various groupings of spectral types (B, dA, dF, dG, dK, and dM).  In some
cases XLFs do not reach the unity since there are upper limits below
the lowest detection so that we have no information below this value.
In general, the XLFs computed over the entire {\em XMM-Newton} FOV tend
to be lower than those computed in the more restricted FOV. Applying
two-sample tests for censored data, we found the difference being
highest in the case of dK stars (confidence level above 99\%), and less
significant (confidence level $\sim$ 93\%) for the dF and dM stars, while
for the B, dA, and dG stars the test is inconclusive. In the case of dF,
dK and, more marginally, dG stars the XLFs computed over the entire FOV
have a lower medians than that derived in the more restricted FOV.

For all the considered spectral types the NGC~2516 {\em XMM-Newton}
XLFs derived considering only the stars in the {\em Chandra} FOV are
statistically indistinguishable from those derived with {\em Chandra}.
In the case of B, dA, and dG stars we have not found statistical
significant difference between the XLFs derived over the entire {\em
XMM-Newton} FOV and those derived from {\em Chandra} data (\cite{H++00}),
while we have found an indication (at 95\% confidence level) of
difference in the cases of dF and dM stars, and a clear difference
(at a confidence level higher than 99.9\%) in the case of dK stars,
but, as we explain below, field star contamination can be a plausible
explanation for these findings.

In comparing the NGC~2516 {\em XMM-Newton} XLFs with those of the
Pleiades (\cite{MSH99}) we have found that, irrespectively of the region
of the {\em XMM-Newton} FOV we are considering, the XLFs of B, dA and dF
stars cannot be distinguished (above the 90\% confidence level), while
the XLFs of dK stars are different at a confidence level higher than
99.9\%. Considering the entire {\em XMM-Netwon} FOV the XLFs of dG and
dM stars are different at $\sim$ 99.9\% level, but, when we limit the
analysis to the {\em Chandra} FOV, the statistical confidence reduces
to $\sim$ 97\% in the case of dG stars, and no difference can be found
for the dM stars.

The NGC~2516 XLFs of dG and dK we have derived over the entire EPIC FOV
reinforce the suggestion based on {\em Chandra} data that NGC~2516 members
are less X-ray luminous than the Pleiades members of analogous spectral
type. This results is particularly strong in the case of dK stars.
Table \ref{tab:XLF}, which compares the median $\log(L_{\rm X})$ values for
NGC~2516 and the Pleiades, indicates that X-ray luminosities are highest
for dF and dG stars.

\section{Summary and Conclusions}

NGC 2516 was chosen as a calibration target in order to ``bore sight''
the alignment of the X-ray telescope with the {\em XMM-Newton} detectors,
however this cluster is of particular scientific interest since, being metal-poor
with respect to the Sun, it allows us to explore the effect of
metallicity on coronal emission level.  Using a 33 ks long EPIC data set
obtained by summing the data, all taken with the thick filter, 
of the three distinct EPIC cameras of two distinct co-pointed observations 
we have reached a limiting sensitivity of $\sim 2.35\times 10^{-15}$ erg~sec$^{-1}$~cm$^{-2}$
and have detected 208 sources. Using only data from a single observation and/or
from the MOS or PN cameras alone we would have reached a factor 2--4 worst sensitivity and
would have detected substantially less sources.
The attained limiting sensitivity 
has allowed us to derive luminosity functions of NGC~2516 members with
a larger number of detected members with respect to the {\em Chandra}
survey (\cite{H++00}).
Together with better photometry made available to us before publication 
(Jeffries, Thurston and Hambly, in preparation), the {\em XMM-Newton}
calibration observations have enabled us
to investigate the late-type population of cluster members and
to explore their coronal emission properties.

dG and dK stars in NGC~2516 tend to be less luminous than the same spectral type
Pleiades members reinforcing the indication coming from the recent analysis of
{\em Chandra} observations (\cite{H++00}). NGC~2516 dM members have the same coronal
emission level of their Pleiades counterparts, confirming the ROSAT HRI-based suggestion
of \cite{MSJ00} that the activity level for dM stars is
insensitive to a change of a factor of two in stellar metallicity. On
the other hand, we find that the median $\log(L_{\rm X})$ value for NGC~2516
dF-type stars when evaluated on the entire EPIC FOV is {\em lower} than that of
the Pleiades, while it has been found {\em higher} than that of the Pleiades analyzing
{\em Chandra} data. Indeed when we restrict the analysis to the dF stars in the
{\em Chandra} survey FOV we recover the {\em Chandra} result. Since the {\em
Chandra} FOV is offset pointed with respect to the {\em XMM-Newton} one, we
do not think that the above findings can be explained by considering 
that we have preferentially included members
surveyed in the lower sensitivity outer part of the {\em XMM-Newton} FOV where we 
would expect a lower
detection percentage. A more viable explanation is that the true members are spatially
concentrated, whereas the contaminating non-members are uniformly distributed. In such
a picture the {\em Chandra} observation is pointed where the cluster members are concentrated
while the present {\em XMM-Newton} survey is pointed in a less central cluster region, 
explaining our finding. However other explanations are possible and further
investigations are required.

Lacking astrometric proper motion studies and extensive radial
velocity surveys of NGC~2516 stars we have been forced to base our study on
photometrically selected members. Such selection is subject
to contamination from field stars that by chance have the same
magnitude and colors as true NGC~2516 members. Based on the very good quality
of the photometry we have used, clearly shown by the CMDs
in Fig.~\ref{fig:CMD}, we think that contamination should not be too severe (but it
could be as large as 15--20\%). However, it is clear that to
be able to draw firmer conclusions regarding metallicity
effects on coronal emission it is crucial that an improved list of
members be obtained complementing available photometric data 
with high precision proper motion studies (unfortunately 
NGC~2516 average motion is only marginally different from the solar reflected one)
and with an extensive campaign of radial velocity measurements.

\begin{acknowledgements}

Based on observations obtained with {\em XMM-New\-ton}, an ESA science mission with instruments
and contributions directly funded by ESA Member States and the USA (NASA).
EPIC was developed by the EPIC Consortium (P.I.,
Dr. M. J. L. Turner). The consortium comprises the 
following Institutes: University of Leicester, University of 
Birmingham, (UK); CEA/Saclay, IAS Orsay, CESR Tou\-louse, (France); 
IAAP Tuebingen, MPE Garching,(Germany); IFC Milan, ITESRE Bologna, 
Osservatorio Astronomico di Palermo, Italy. EPIC is funded by: PPARC, CNES, DLR and ASI.
SS, GM, FD and EF acknowledge partial support by Agenzia Spaziale Italiana and 
Ministero della Ricerca Scientifica e Tecnologica.
NG is supported by the European Union (Marie Curie Individual grant;
HPMF-CT-1999-00228). KRB and JPP acknowledge the financial
support of the UK Particle Physics and Astronomy Research Council.
\end{acknowledgements}

\end{document}